\documentclass[%
reprint,
 amsmath,amssymb,
 aps,
prb,
]{revtex4-2}

\usepackage{graphicx}
\usepackage{dcolumn}
\usepackage{bm}
\usepackage{color}
\usepackage{colortbl, array, xcolor}

\begin{document}
\newcolumntype{A}{>{\columncolor{yellow!30}}c}
\preprint{APS/123-QED}

\title{Systematic study of superconductivity in few-layer $T_{\rm d}$-MoTe$_2$}

\author{T. Wakamura,$^1$ M. Hashisaka,$^2$ Y. Nomura$^{3, 4}$, M. Bard$^1$, S. Okazaki,$^5$ T. Sasagawa,$^{5, 6}$ T. Taniguchi,$^7$ K. Watanabe,$^8$ K. Muraki,$^1$}
\author{N. Kumada$^1$}
\affiliation{$^1$Basic Research Laboratories, NTT, Inc., 3-1 Morinosato-Wakamiya, Atsugi, 243-0198, Japan}
\affiliation{$^2$Institute for Solid State Physics, the University of Tokyo, 5-1-5 Kashiwa-no-ha, Kashiwa, 277-8581, Japan}
\affiliation{$^3$Institute for Materials Research (IMR), Tohoku University, 2-1-1 Katahira, Aoba-ku, Sendai, 980-8577, Japan}
\affiliation{$^4$Advanced Institute for Materials Research (WPI-AIMR), Tohoku University, 2-1-1 Katahira, Aoba-ku, Sendai, 980-8577, Japan}
\affiliation{$^5$Materials and Structures Laboratory, Institute of Science Tokyo, 4259 Nagatsuta, Yokohama, 226-8501, Japan}
\affiliation{$^6$Research Center for Autonomous Systems Materialogy, Institute of Science Tokyo, 4259 Nagatsuta, Yokohama, 226-8501, Japan}

\affiliation{$^7$Research Center for Materials Nanoarchitectronics, National Institute for Materials and Science, 1-1 Namiki, Tsukuba, 305-0044, Japan}

\affiliation{$^8$Research Center for Electronic and Optical Materials, National Institute for Materials and Science, 1-1 Namiki, Tsukuba, 305-0044, Japan}%

\date{\today}

\begin{abstract}
\textcolor{black}{We present a systematic investigation of superconductivity in a topological superconductor candidate $T_{\rm d}$-MoTe$_2$ in the few-layer limit. By examining multiple mechanically exfoliated samples with different thicknesses, substrates and crystal qualities, we quantitatively correlate superconducting temperature ($T_c$) with disorder, carrier density, carrier type and mobility. By integrating these experimental findings with first-principles calculations, we reveal the relationship between the band structure and superconductivity in this material. Notably, in 2 L samples we access a highly hole-doped regime that has not been systematically explored in previous experiments, providing a complementary perspective to earlier studies. In this regime, we demonstrate that superconductivity can be realized in a manner consistent with a conventional phonon-mediated $s_{(++)}$-wave pairing.} 
\end{abstract}

\maketitle




\section{Introduction}
A wide variery of electronic properties and their coexistence stimulate recent growing interest in two-dimensional (2D) materials. $T_{\rm d}$-MoTe$_2$, a semimetallic transition metal dichalcogenide (TMD), is known in its bulk form as a type-II Weyl semimetal \cite{MoTe2_Weyl1, MoTe2_Weyl2}. Notably, it undergoes a superconducting transition at approximately 100 mK \cite{Qi}, making it a promising candidate for topological superconductivity \cite{MR2, Huang_2024, Li_2024}.

The superconducting properties of $T_{\rm d}$-MoTe$_2$ are highly intriguing in their own right, even without considering its potential topological nature. For instance, the possibility of unconventional multi-band $s_\pm$-wave pairing is proposed both for bulk and thin layers in previous studies \cite{Guguchia, Jindal}. Recent reports highlight an anomalous enhancement of the superconducting critical temperature ($T_c$) with decreasing thickness, the origin of which remains unclear \cite{MoTe2cn, Rhodes}. The crystal structure of $T_{\rm d}$-MoTe$_2$, lacking inversion symmetry, enables nonreciprocal superconducting transport \cite{Wakamura_2024, Du_2024, Zhou_2025}.

While these results demonstrate that $T_{\rm d}$-MoTe$_2$ hosts a variety of exotic superconducting phenomena, their detailed characteristics and mechanisms remain elusive. For instance, a previous study reports that $T_c$ was insensitive to incresing disorder in bulk $T_{\rm d}$-MoTe$_2$, \textcolor{black}{indicating the possibility of the conventional $s_{(++)}$-wave rather than the unconventional $s_\pm$-wave pairing \cite{Piva2023}}. 
The enhancement of $T_c$ has been reported exclusively in mechanically exfoliated samples, whereas samples grown via chemical vapor deposition (CVD) exhibit a suppression of $T_c$ with decreasing thickness \cite{Cui_2019}. {\color{black}Even among mechanically exfoliated samples, however, there are notable discrepancies in the previously reported results. Although superconductivity with $T_c \sim$ 7 K has been observed in monolayers near the charge neutrality point (CNP) \cite{Rhodes}, another study reported insulating behavior under similar conditions \cite{Tang_2023}. A critical issue in previous studies is that the quality of MoTe$_2$ crystals used for exfoliation varies, and $T_{\rm d}$-MoTe$_2$ flakes were measured on different substrates, such as SiO$_2$ or hexagonal boron-nitride (hBN). To elucidate superconductivity and its gap symmetry of thin-layer $T_{\rm d}$-MoTe$_2$, it is essential to consider the effects of substrate type, disorder and carrier density on superconductors. However, most of them have not yet been systematically explored.}

In this paper, we present a systematic study of superconductivity in mechanically exfoliated $T_{\rm d}$-MoTe$_2$ thin layers, using multiple samples with thicknesses ranging from 2 to 23 layers (Ls). For \textcolor{black}{2 L and 4 L, the latter of which in particular has been poorly explored}, we measure several samples with different types of substrates, and discuss variations among them. Gate-dependent transport measurements in both the normal and superconducting regimes, combined with normal-state magnetoresistance measurements, reveal correlations between $T_c$ and electronic transport parameters, such as carrier density and mobility. By integrating these experimental observations with first-principles calculations, we investigate the possible pairing symmetry. Our analysis suggests that multiple bands are not required for superconductivity \textcolor{black}{at least in the hole-doped regime, and conventional $s_{(++)}$-wave superconductivity is possible}. These findings provide critical insights into the intrinsic superconducting behavior of $T_{\rm d}$-MoTe$_2$ and offer valuable guidance for realizing topological superconductivity in this material.

This paper is organized as follows. Section II outlines the experimental methods. In Section III, we present the thickness dependence of $T_c$ for $T_{\rm d}$-MoTe$_2$ in the thin limit. Influence of electronic transport parameters such as carrier densities and mobilities on $T_c$ is discussed in Section IV among different samples and also in a single sample for certain thicknesses. Section V compares the experimental results with the density of states and band structures obtained from first-principles calculations, and discusses the possible pairing symmetry. Finally, Section VI summarizes our findings and outlines future directions for exploring exotic superconductivity in $T_{\rm d}$-MoTe$_2$. 


\section{Methods}

We prepared our samples from high-quality $T_{\rm d}$-MoTe$_2$ crystals grown by the flux method \cite{Wang2021}, with a residual-resistivity ratio (RRR) of $\sim$ 1,000, a measure of the crystal quality defined as the ratio of the resistance at 4 K to that at room temperature. Thin flakes were obtained by mechanical exfoliation and they were subsequently transferred onto pre-patterned electrodes on a Si/SiO$_2$ chip or hexagonal boron-nitride (h-BN) deposited on a Si/SiO$_2$ chip using a typical dry transfer technique with polycarbonate (PC) and poli-dimethylpolysiloxane (PDMS) \cite{PC} in an argon-filled glovebox with a low concentration of O$_2$ and H$_2$O ($<$ 0.5 ppm) to avoid degradation of $T_{\rm d}$-MoTe$_2$.  The thickness of the flakes was first identified by the optical contrast under the microscope and then confirmed by atomic force microscopy (AFM) after the transport measurements.  
Metal electrodes were fabricated by electron-beam lithography and electron-beam evaporation of thin Pt (or Au) and Ti. 
Electrical measurements were performed with a lock-in amplifier using a $^3$He low-temperature measurement system. The magnetic field was applied perpendicular to the sample plane. More experimental details are given in \cite{Wakamura_2024}.

\begin{figure}[tb!]
\begin{center}
\includegraphics[width=6.5cm,clip]{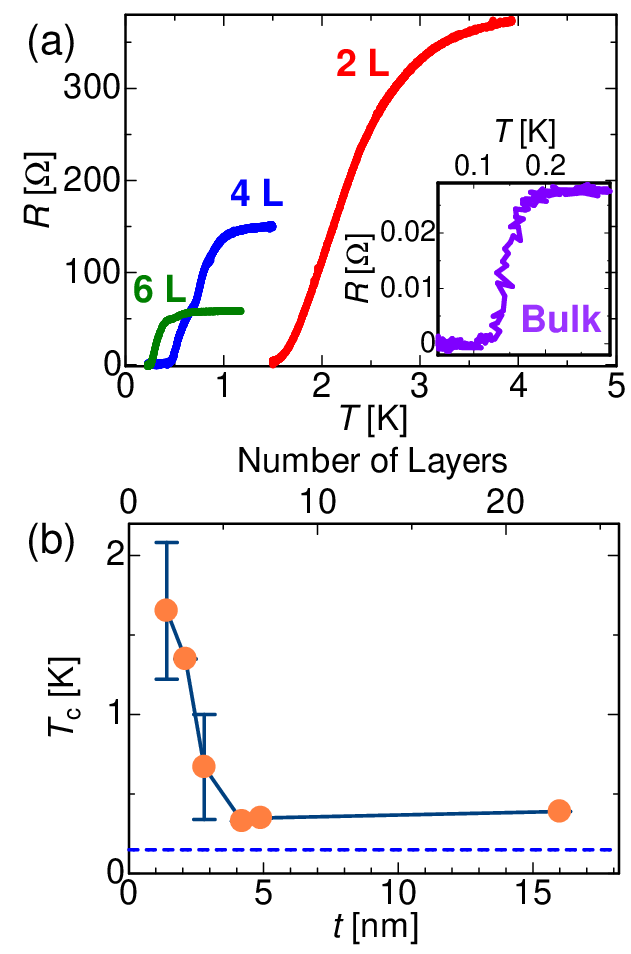}
\caption{(a) {\color{black} Typical t}emperature ($T$) dependence of resistance ($R$) for samples with different numbers of layers. Inset: $T$-$R$ from a bulk sample for reference, where $T_c$ = 150 mK.
(b) Thickness dependence of the superconducting critical temperature ($T_c$), showing a rapid increase with decreasing thickness, particularly for $t <$ 5 nm (approximately 7 L). The error bars represent the variation among samples. The horizontal dashed line indicates $T_c \sim$ 150 mK for the bulk sample with $t >$ 100 nm.}
\label{fig1}
\end{center}
\end{figure}

\begin{table*}[tb!]
\caption{\label{tab:table3}Summary of the number of layers, the substrate type, the residual resistivity ratio (RRR), and $T_c$ that we have measured. "-" in the row of $T_c$ expresses no signatures of superconductivity at least down to 230 mK. "-" in the row of $n_e$ and $n_h$ represents that the carrier densities could not be obtained because the sample was broken before performing magnetotransport measurements.}
\begin{ruledtabular}
\begin{tabular}{ccccccc}
 Sample Name & Number of Layers & $n_e$($V_g$=0) [$\times10^{13}$cm$^{-2}$] & $n_h$($V_g$=0) [$\times10^{13}$cm$^{-2}$] & Substrate & RRR & $T_c$ [K] \\ \hline
 MTS4 & 2 & 2.0$\times 10^{-2}$ & 2.0 & hBN & 3.7 & 2.2  \\
 MTS10 & 2 & 6.5$\times 10^{-3}$ & 1.2 & hBN & 1.7 & 1.2  \\
 MTS18 & 2 & 1.6$\times 10^{-2}$ & 1.4 & SiO$_2$ & 3.3 & 1.6 \\
 MTS17 & 3 & 3.0$\times 10^{-2}$ & 9.1 & SiO$_2$ & 4.0 & 1.4 \\
 MTS3 & 4 & 4.9 & 2.7 & SiO$_2$ & 4.6 & 0.75  \\
 MTS5 & 4 & 6.7 & 0.39 & hBN & 11 & 0.19\footnotemark[1] \\
 MTS7 & 4 & - & - & hBN & 2.9 & -  \\
 MTS9 & 4 & - & - & hBN & 14 & 1.2  \\
 MTS14 & 4 & 2.4 & 2.6 & SiO$_2$ & 23 & - \\
 MTS19 & 4 & 1.5 & 1.7 & SiO$_2$ & 14 & 0.5 \\
 MTS6 & 6 & 5.7 & 0.24 & hBN & 3.3 & 0.33  \\
 MTS8 & 7 & 3.7 & 0.80 & hBN & 23 & 0.35 
\label{Table1}
\end{tabular}
\end{ruledtabular}
\footnotetext[1]{\textcolor{black}{Using a partial drop of resistance observed above 230 mK, we estimate $T_c$ by fitting the temperature dependence of resistance with the Halperin-Nelson equation.}}
\end{table*}


\section{Thickness dependence of $T_c$ in thin samples}

To evaluate the superconducting properties of our few-layer $T_{\rm d}$-MoTe$_2$ samples, we first measured the temperature dependence of resistance ($R$) using multiple samples with varying thicknesses. Representative $R–T$ curves are shown in Fig. \ref{fig1}(a), where the superconducting transition clearly shifts to higher temperatures as the thickness decreases. The $T_c$ for the 2 L sample ($\sim$ 2.2 K) in Fig. \ref{fig1}(a) is consistent with previously reported values \cite{Rhodes, Jindal}. The inset in Fig. \ref{fig1}(a) presents the $R-T$ curve taken from our bulk ($t >$  100 nm) sample and $T_c \sim$ 150 mK, slightly higher than the reported value of $T_c$ for bulk $T_{\rm d}$-MoTe$_2$ ($T_c \sim$ 100 mK) \cite{Qi}. Here, we define $T_c$ as the temperature at which $R$ reaches half of its normal-state value, taken just above the onset of the superconducting transition. The thickness dependence of $T_c$ is plotted in Fig. \ref{fig1}(b) over a broad range of thicknesses $t$. A rapid increase in $T_c$ is observed particularly below 7 L ($\sim$ 5 nm). The horizontal dashed line in Fig. \ref{fig1}(b) represents $T_c$ for our bulk sample ($T_c \sim$ 150 mK). Notably, even the sample with $t$ = 16 nm ($\sim$ 20 L) shows an enhanced $T_c$ compared to the bulk sample. This observation indicates that the deviation from the bulk regime may already be effective even for samples with 20 L thickness. 

Comparison between $T_c$ and other parameters for multiple samples are shown in Table I. Interestingly, we did not observe a systematic correlation between the substrate type \textcolor{black}{(hBN or SiO$_2$)} and $T_c$. \textcolor{black}{This suggests that, in contrast to graphene where substrate effects are pronounced, $T_c$ of thin $T_{\rm d}$-MoTe$_2$ is primarily governed by properties of $T_{\rm d}$-MoTe$_2$ themselves, and substrate-related influences are relatively subtle in our devices.} Below, we discuss effects of electronic transport parameters in more detail.

\begin{figure*}[tb!]
\begin{center}
\includegraphics[width=14.5cm,clip]{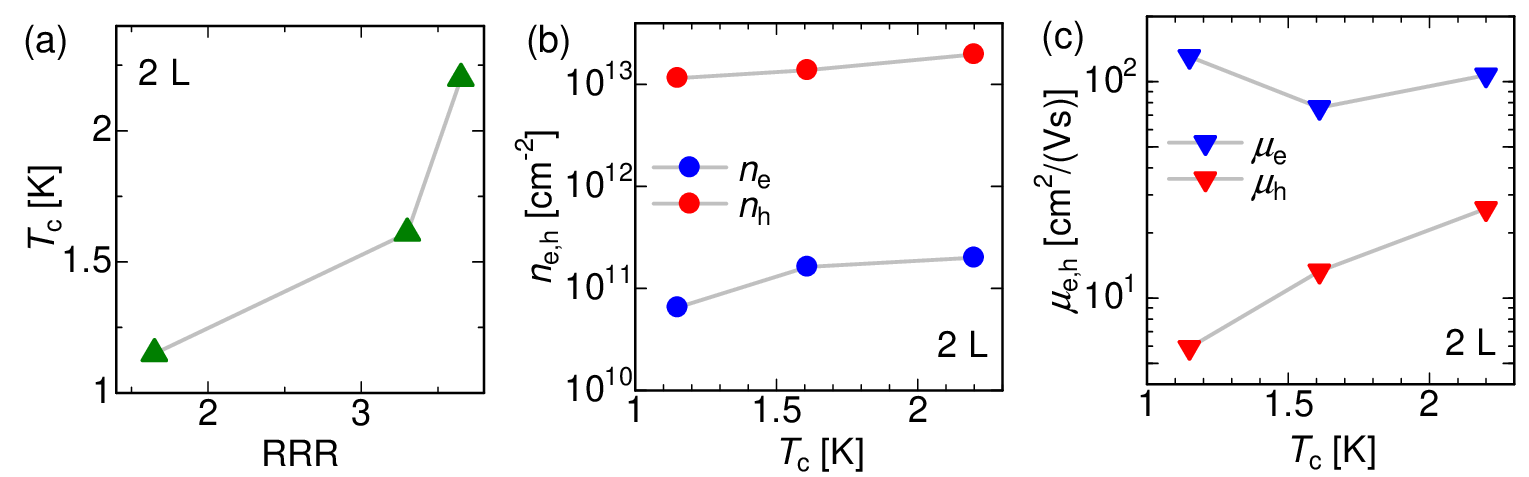}
\caption{(a) $T_c$ as a function of the residual resistivity ratio (RRR) for three different 2 L samples \textcolor{black}{(MTS10, MTS18 and MTS4 from left to right, the same for (b) and (c))}. (b) Electron ($n_e$) and hole ($n_h$) carrier density dependence of $T_c$ from different 2 L samples. (c) Electron ($\mu_e$) and hole ($\mu_h$) mobility dependence of $T_c$ from different 2 L samples. }
\label{fig2L_sample_dep}
\end{center}
\end{figure*} 

\section{Influence of electronic transport parameters on $T_c$}
\subsection{2 L case}
\subsubsection{Comparison among different samples: Relation between RRR, carrier density, mobility and $T_c$}
We next compare different samples with the same thickness to examine the influence of electronic transport parameters on superconductivity, focusing first on the 2 L samples. Figure \ref{fig2L_sample_dep}(a) displays the relationship between RRR and $T_c$ for the 2 L samples. 

In previous studies on bulk $T_{\rm d}$-MoTe$_2$, a suppression of $T_c$ with decreasing RRR was reported, suggesting that superconductivity is sensitive to disorder; an observation considered indicative of unconventional pairing mechanisms \cite{Rhodes_2017, Guguchia}. In our 2 L samples, as shown in Fig. \ref{fig2L_sample_dep}(a), $T_c$ decreases monotonically with decreasing RRR, consistent with the trend observed in bulk. A summary of $T_c$ and RRR values for each sample is provided in Table I, including those from the sample with different thicknesses.

The robustness of $s_\pm$-wave superconductivity against disorder has been extensively discussed in the context of iron-based superconductors \cite{Onari_2009}. According to Anderson's theorem, superconducting states other than conventional $s$-wave pairing are generally fragile in the presence of disorder \cite{Anderson1959}. The observed suppression of $T_c$ with decreasing RRR is consistent with theoretical predictions for $s_\pm$-wave superconductivity \cite{Onari_2009}. Experimentally, however, a reduction in $T_c$ with increasing disorder has also been reported in conventional $s$-wave superconductors \cite{Mondal_2011}, and is therefore not an exclusive indicator of unconventional pairing symmetry. Thus, additional experimental signatures are necessary to conclusively determine the superconducting pairing symmetry in $T_{\rm d}$-MoTe$_2$. 

In addition to RRR, other electronic transport parameters also influence superconductivity. We performed normal-state magnetoresistance measurements to estimate the carrier density and mobility of the samples. Owing to the semimetallic nature of $T_{\rm d}$-MoTe$_2$, both electrons and holes must be considered in the transport analysis. Details of the measurement procedures and analysis methods are provided in Appendix A. Figures \ref{fig2L_sample_dep}(b) and (c) summarize the correlations between $T_c$ and carrier density or mobility. Notably, all 2 L samples are hole-doped [see Fig. \ref{fig2L_sample_dep}(b)], with the electron contribution being negligible, and $T_c$ increases with increasing carrier density. Regarding carrier mobilities, mobilities of holes, the dominant carrier type mediating superconductivity, are lower than those of electrons. On the other hand, when focused on $\mu_h$ among different samples, larger $\mu_h$ is advantageous for higher $T_c$, consistent with the relation between RRR and $T_c$ shown in Fig. \ref{fig2L_sample_dep}(a).

\subsubsection{Effects of gate-tuned carrier density within a single sample}

In the previous part, we examined the relationships between $T_c$ and several electronic transport parameters across different samples with 2 L thickness. While such inter-sample comparisons are commonly used for bulk crystals, superconductivity is generally influenced by multiple intertwined factors, making it difficult to isolate the effect of a single parameter. A major advantage for thin metallic 2D materials is that their transport properties can be modulated by a gate voltage using a standard solid-state gate, owing to the sample thickness comparable to or smaller than the Thomas-Fermi (TF) screening length \cite{Gate2, Chiral, formula}.
In the single carrier case, the TF screening length  $\lambda_{\rm TF}$ can be estimated using the formula $\lambda_{\rm TF} =  \sqrt{(\epsilon_0 \hbar^2 \pi^{4/3})/(3^{1/3} m^* q^2 n^{1/3})}$, where $m^*$, $q$, and $n$ are the effective mass, charge of carriers, and carrier concentration, respectively. 
Employing an isotropic three-dimensional model for simplicity, and assuming $m^* \sim$ 0.3$m_0$ (where $m_0$ is the free electron mass) averaged over the three crystal axes \cite{Rhodes}, $q = |e|$, and the carrier density of the dominant carrier type, we obtain $\lambda_{\rm TF}$ on the same order as the lattice constant of the $c$-axis (= 0.7 nm). 

\begin{figure*}[tb!]
\begin{center}
\includegraphics[width=16.5cm,clip]{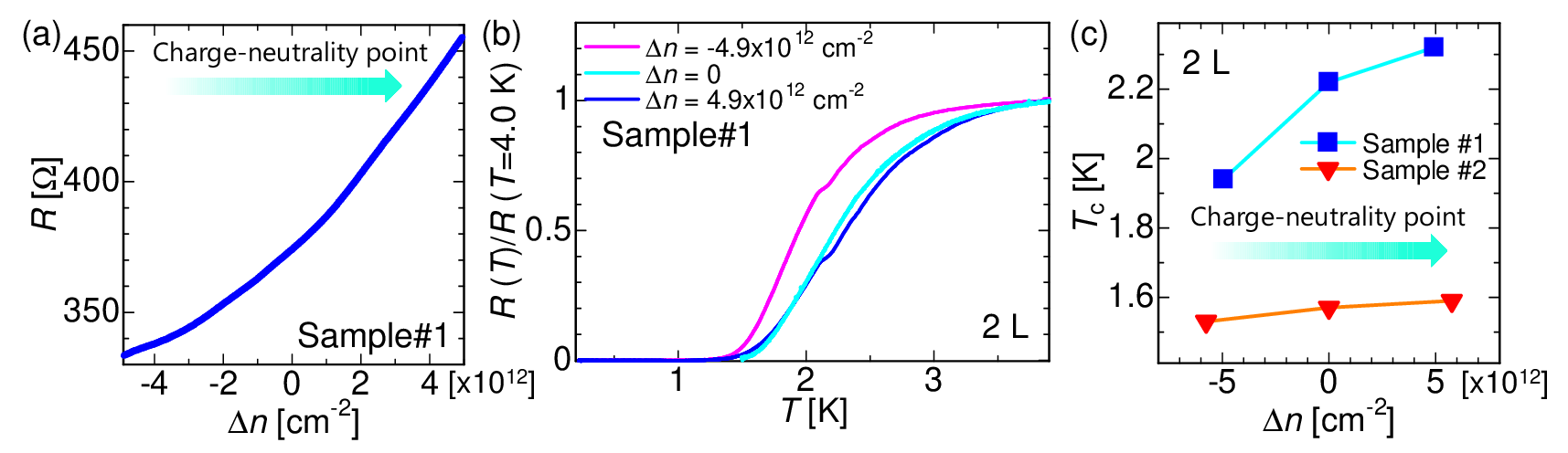}
\caption{(a) Gate voltage dependence of resistance in the normal state for a 2 L sample (MTS4). The gate voltage ($V_g$) is converted to $\Delta n = C_gV_g$ with the gate capacitance $C_g$. (b) Temperature dependence of resistance for the 2 L sample, normalized by the value at 4.0 K with different $V_g$. (c) $T_c$ as a function of $\Delta n$ for different 2 L samples (Sample\#1: MTS4, Sample\#2: MTS18), showing the monotonic increase of $T_c$ as the Fermi level moves toward the CNP. The CNP is located in the positive side beyond the range of $x$-axis in (a) and (c) (see arrows).}
\label{fig2L_gate}
\end{center}
\end{figure*} 

The gate voltage ($V_g$) dependence of the normal-state resistance provides a useful means to cross-check the dominant carrier type in the system. Furthermore, by measuring the $R–T$ curves under various $V_g$, we gain insight into the carrier and Fermi-level dependence of $T_c$. Figure \ref{fig2L_gate}(a) shows the $V_g$ dependence of the normal-state resistance for a 2 L sample. Here, $\Delta n_g = C_g V_g$, where $C_g = \varepsilon_r \varepsilon_0/(ed)$ is the gate capacitance per unit area with a relative permittivity $\varepsilon_r$, the permittivity of vacuum $\varepsilon_0$, the elementary charge $e$ and the thickness of the gate insulator $d$. $C_g$ = 6.2$\times$10$^{11}$ cm$^{-2}$V$^{-1}$ and 7.2$\times$10$^{10}$ cm$^{-2}$V$^{-1}$ for Sample\#1 and Sample\#2, respectively. The resistance increases as we drive $V_g$ from the negative to positive side, indicating that the sample is originally hole-doped. This is consistent with the dominant carrier type estimated from the magnetoresistance measurements, \textcolor{black}{$n_e=2.0\times10^{11}$ cm$^2$ and $n_h=2.0\times10^{13}$ cm$^2$ for Sample\#1, and $n_e=1.6\times10^{11}$ cm$^2$ and $n_h=1.4\times10^{13}$ cm$^2$ for Sample\#2 at $V_g$=0 V}. Figure \ref{fig2L_gate}(b) presents the $R–T$ curves at various $V_g$ for a 2 L sample, and the extracted $T_c$ values as a function of $V_g$ are summarized in Fig. \ref{fig2L_gate}(c). We find that $T_c$ increases monotonically as the Fermi level approches the CNP for different samples. Note that this result is not consistent with the trend shown in Fig. \ref{fig2L_sample_dep}(b). \textcolor{black}{However, because not only carrier densities but also RRR (or mobility) vary among different samples, the variation of $T_c$ in Fig. \ref{fig2L_sample_dep}(b) may be induced by the mobility difference. To investigate the effect of carrier density on $T_c$ uniquely, the $V_g$ dependence in a single sample provides more direct information.}

\subsection{4 L case}

\begin{figure*}[tb!]
\begin{center}
\includegraphics[width=16.5cm,clip]{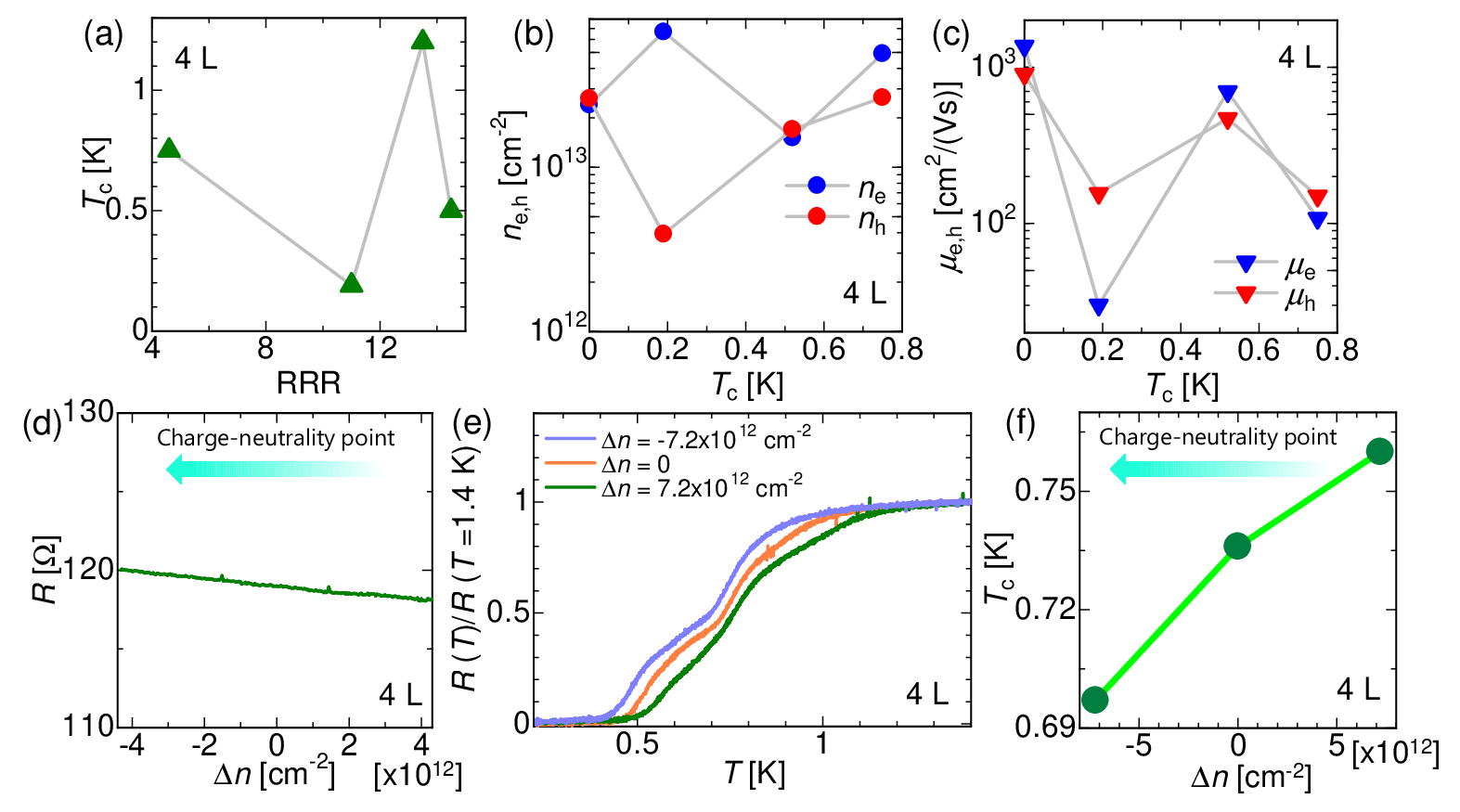}
\caption{(a) RRR for four different 4 L samples (MTS3, MTS5, MTS9 and MTS19). (b) Electron ($n_e$) and hole ($n_h$) carrier density dependence of $T_c$ from different 4 L samples (MTS14, MTS5, MTS19 and MTS3). (c) Electron ($\mu_e$) and hole ($\mu_h$) mobility dependence of $T_c$ from different 4 L samples. (d) Gate dependence of resistance in the normal state for a 4 L sample (MTS3). (e) Temperature dependence of resistance for the 4 L sample (MTS3), normalized by the value at 1.4 K, taken at different $V_g$. (f) $T_c$ plotted as a function of $\Delta n$ for the 4 L sample (MTS3). Similarly to the 2 L case, $T_c$ monotonically increases with electron doping. The CNP is located in the negative side beyond the range of $x$-axis in (d) and (f) (see arrows).}
\label{fig4L}
\end{center}
\end{figure*}

We also investigate relations between the electronic transport parameters and superconductivity for 4 L samples. Results are summarized in Figs. \ref{fig4L}(a)-(c). In comparison to the 2 L case, no clear trends are found between RRR and $T_c$ [Fig. \ref{fig4L}(a)]. Surprisingly, some cleaner samples with higher RRR values do not show any signatures of superconductivity down to 230 mK, while several more disordered samples exhibit finite $T_c$ ($>$ 230 mK). Such a non-monotonic dependence of $T_c$ on RRR observed in the 4 L samples remains difficult to interpret and cannot be readily explained within the existing theoretical frameworks. More details are given in Table I including information for samples which do not show superconducting transition.


Normal-state magnetoresistance measurements reveal no clear tendency between carrier densities or mobilities and $T_c$ [see Figs. \ref{fig4L}(b) and (c)]. These results indicate that it is difficult to draw some conclusion about the effects of the electronic transport parameters on $T_c$ by comparing different samples with the same 4 L thickness.

Because of larger thickness, variations of the normal state resistance and $T_c$ by the standard electrostatic gating are smaller than those for the 2 L case. For some 4 L samples, indeed, we do not observe a clear variation of the normal state resistance. Nevertheless, from a particular sample we can obtain the dependence of $T_c$ on carrier densities and dominant carrier type. Figure \ref{fig4L}(d) presents such an example, where the gate dependence of normal-state resistance shows that the sample is electron-doped. \textcolor{black}{This is consistent with the carrier densities estimated from the normal state magnetoresistance measurements, $n_e=4.9\times10^{13}$ cm$^2$ and $n_h=2.7\times10^{13}$ cm$^2$ at $V_g$=0 V.} $R-T$ curves at different $V_g$ values shown in Fig. \ref{fig4L}(e) demonstrate that electron doping enhances $T_c$, as plotted in Fig. \ref{fig4L}(f).

\section{Comparison with first-principles calculation}

\begin{figure*}[tb!]
\begin{center}
\includegraphics[width=15cm,clip]{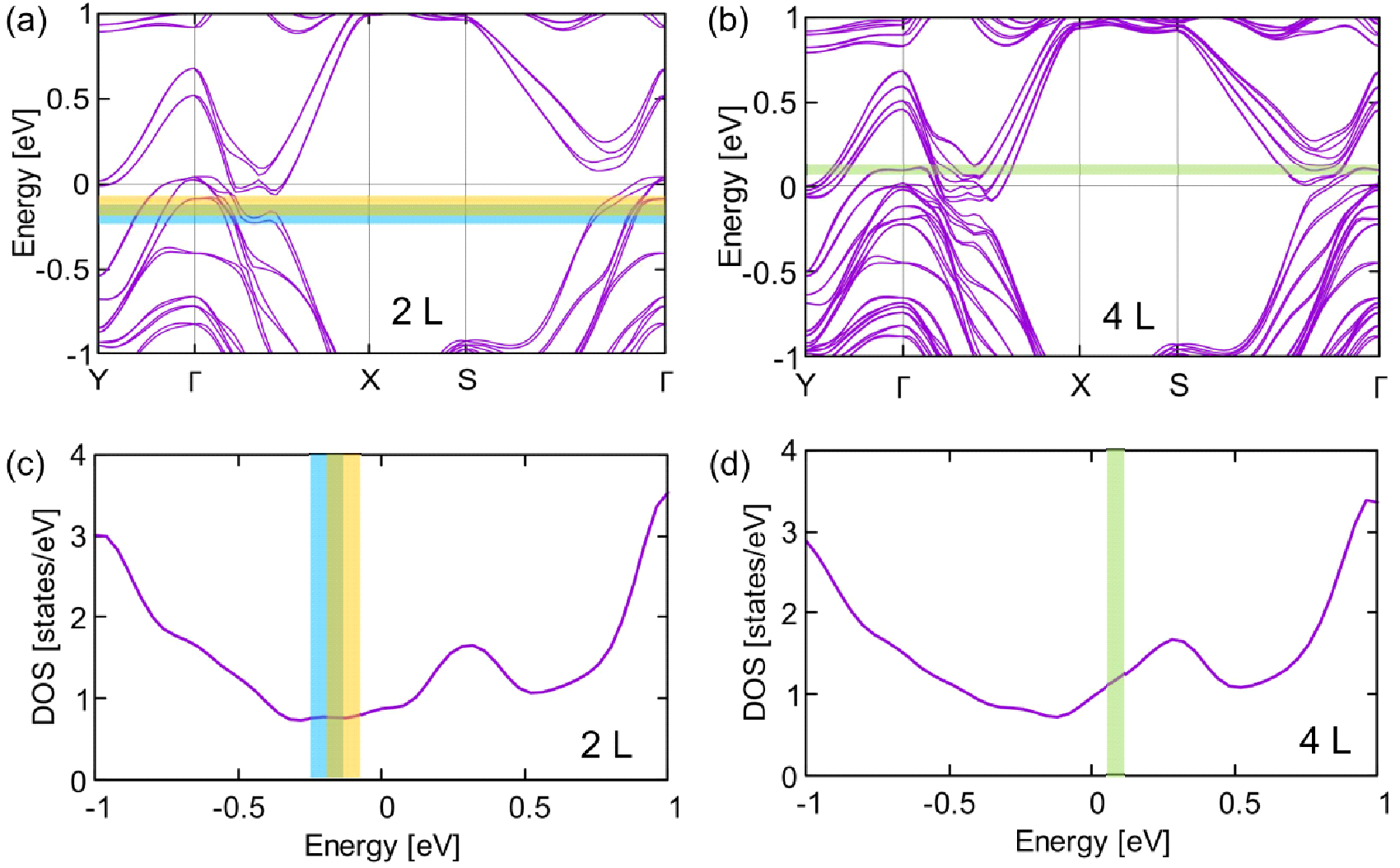}
\caption{(a) and (b) Band structures for the 2 L and 4 L obtained from first-principles calculations. (c) Density of states (DOS) per formula unit for the 2 L estimated from (a). (d) Similar DOS per formula unit shown for the 4 L. The light-blue-shaded (orange-shaded) rectangles in (a) and (b) express the energy range where the Fermi level moves by modulating the gate voltage for Sample\#1 (Sample\#2, see Figs. 3) for the 2 L samples. The light-green-shaded rectangle in (c) and (d) is similar for one of the 4 L samples.}
\label{fig_DFT}
\end{center}
\end{figure*}

The experimental results presented above can be used to infer the superconducting pairing symmetry of $T_{\rm d}$-MoTe$_2$, particularly when combined with first-principles calculations. These calculations provide the band structures and corresponding density of states (DOS) for $T_{\rm d}$-MoTe$_2$ at various layer thicknesses. The calculated band structures and DOS for the 2 L and 4 L cases are shown in Figs. \ref{fig_DFT}(a) and (b), and Figs. \ref{fig_DFT}(c) and (d), respectively. In an ideal single crystal, the Fermi level lies at the CNP, where both electron and hole pockets coexist and are perfectly compensated, resulting in a charge-neutral system. Experimentally, however, the Fermi level is often shifted away from the CNP, as observed in the gate-voltage dependence of resistance discussed in Section IV. Using the difference in carrier densities between electrons and holes obtained from magnetoresistance measurements and the DOS assuming parabolic bands for electron and hole pockets, we can estimate the shift of the Fermi level relative to the CNP. In Figs. \ref{fig_DFT}(a)–(d), we highlight the energy ranges corresponding to the Fermi level variation induced by gate voltage for different samples.

With the estimated Fermi level positions, we can now discuss the implications for the superconducting pairing symmetry in few-layer $T_{\rm d}$-MoTe$_2$. Since the electronic states near the Fermi level are primarily responsible for superconductivity, identifying their character is essential for understanding the pairing mechanism. By comparing the experimentally observed $T_c$ evolution with carrier density (or equivalently, the gate voltage) to the calculated DOS and band dispersion, we can assess whether the observed superconductivity is consistent with a conventional $s_{(++)}$-wave or an unconventional $s_\pm$-wave gap structure.

The $s_\pm$-wave pairing, proposed as a candidate for unconventional superconductivity in $T_{\rm d}$-MoTe$_2$ \cite{Guguchia, Jindal}, is typically mediated by antiferromagnetic fluctuations that promote multiband Cooper pairing involving both electron and hole pockets \cite{Chubukov2015, Fernandes2022, Bang2017, Stewart2011}. A previous study identified a peak in the Lindhard susceptibility at a wave vector connecting an electron pocket and a hole pocket as a possible hallmark of $s_\pm$-wave pairing \cite{Jindal}. This pairing mechanism relies critically on the multiband nature of the Fermi surface, particularly the coexistence of both electron and hole pockets \cite{Kuroki2008, Mazin2008}.

\textcolor{black}{An important outcome of our study is the observation of superconductivity in the highly hole-doped regime of 2 L samples with $T_c$ comparable to earlier works. In the two 2 L samples investigated here, the Fermi level lies entirely within the hole-doped region. Even under electrostatic gating, the accessible carrier density range (light-blue and orange regions in Fig. \ref{fig_DFT}(a) for Sample\#1 and Sample\#2, respectively) remains on the hole-doped side, below the energy range where the hole and electron pockets coexist.
This observation has direct implications for the superconducting pairing symmetry. The multiband $s_\pm$-wave pairing scenario proposed in earlier studies relies on interband pairing between electron and hole pockets. Our results demonstrate that superconductivity can be sustained in the absense of electron pockets, indicating that multiband pairing is not prerequisite for superconductivity in 2 L $T_{\rm d}$-MoTe$_2$, at least in the hole-doped regime explored here.} The 4 L case is shown in Fig. \ref{fig_DFT}(b) with a light-green-shaded rectangle. Note here that because only single sample is available which exhibits a clear gate modulation of the normal-state resistance and $T_c$, we use the data from this sample. Both electron and hole pockets are present at the Fermi level in part of the accessible energy range for the 4 L case. This suggests that multiband contributions may still play a role in 4 L samples.


Based on the single-band nature in the hole-doped region discussed above, we then assume a conventional $s_{(++)}$-wave pairing, rather than the unconventional $s_\pm$-wave, to explain our experimental observations on the carrier density dependence of $T_c$. In the BCS theory, $T_c$ is given by the relation \cite{Tinkham2}
\begin{equation}
k_{\rm B} T_c = 1.13 \hbar \omega_{\rm D} \exp \left( -\frac{1}{N(0) V} \right),
\label{BCS} 
\end{equation}
where $\omega_{\rm D}$, $N(0)$, and $V$ are the Debye frequency, the DOS at the Fermi level, and the electron-phonon coupling potential, respectively. This formula assumes three-dimensional (3D) superconductors, but is also approximately applicable to 2D superconductors, as demonstrated in previous studies of superconductivity in TMDs \cite{Yang_2018, Lee_2021}.   
Figure \ref{fig_DFT}(c) presents the DOS in the Fermi level energy region in the 2 L case. In this energy region, the DOS is almost unchanged for both samples, indicating that there is littile contribution of the DOS to the increase of $T_c$ as the Fermi level moves toward the CNP. 

A recent theoretical study on monolayer $T_{\rm d}$-MoTe$_2$ reports the carrier doping dependence of the electron-phonon coupling constant $\lambda = N(0)V$, which may help explain our experimental findings \cite{Lee_2021}. The calculation shows that $\lambda$ increases monotonically from the hole-doped to the electron-doped regime. As a result, $T_c$ also increases with electron doping, up to the highly electron-doped region where superconductivity is suppressed due to the emergence of charge-density wave (CDW) order. Although the phonon spectra may differ slightly between monolayer and bilayer (2 L) $T_{\rm d}$-MoTe$_2$, a similar monotonic increase in $\lambda$ with electron doping is expected to hold for the 2 L case. This trend is consistent with our experimental observation of enhanced $T_c$ as the Fermi level approches the CNP.

A similar mechanism may also apply to the 4 L case, as shown in Figs. \ref{fig_DFT}(b) and \ref{fig_DFT}(d). In this case, however, in addition to the variation in the electron-phonon coupling constant $\lambda$ with electron doping, the density of states (DOS) in the relevant energy range [Fig. \ref{fig_DFT}(d)] also changes significantly. Therefore, the observed increase in $T_c$ can be attributed to the enhancement of $V$, the increase in DOS, or a cooperative effect of both. In both the 2 L and 4 L cases, the variation of $T_c$ with carrier density is well accounted for by a conventional $s_{(++)}$-wave superconducting state mediated by electron-phonon coupling.

\textcolor{black}{We stress here that our results do not rule out the possibility of other pairing mechanisms under different experimental conditions. In particular, the dual-gated bilayer devices studied in \cite{Jindal} access a much wider doping range and involve an additional displacement field, which could induce ferroelectricity and modify the pairing interaction. Our work therefore provides a complementary perspective, demonstrating that in single-gated devices and in the heavily hole-doped regime, superconductivity in thin $T_{\rm d}$-MoTe$_2$ is consistent with a conventional phonon-mediated $s_{(++)}$-wave pairing.} 

\section{Conclusions and Future outlook}
In this study, we investigated superconductivity in few-layer $T_{\rm d}$-MoTe$_2$ using multiple samples with varying thicknesses. We focused in particular on the 2 L and 4 L samples, and examined the effects of disorder, carrier density, mobility \textcolor{black}{and substrates on $T_c$ using multiple devices and via the gate-voltage dependence of normal-state resistance and $T_c$ as well as normal-state magnetoresistance measurements.} In addition, we performed first-principles calculations and compared the theoretical results with our experimental observations to gain insight into the superconducting pairing symmetry. 
We found that superconductivity can be realized in the heavily hole-doped regime without electron contribution, consistent with $s_{(++)}$-wave pairing symmetry. This conclusion is further supported by a recent theoretical study on the doping dependence of the electron-phonon coupling constant in monolayer $T_{\rm d}$-MoTe$_2$.



While we successfully measured superconductivity in multiple few-layer $T_{\rm d}$-MoTe$_2$ samples, the origin of one of its most intriguing properties, namely the significant enhancement of $T_c$ with decreasing thickness, remains unresolved. A similar $T_c$ enhancement has been reported in the transition metal dichalcogenide TaS$_2$ as the number of layers is reduced \cite{Navarro-Moratalla2016, Yang_2018}, and it is driven by the suppression of a charge density wave (CDW) order. In contrast to TaS$_2$, no experimental observation of CDW order has been reported for $T_{\rm d}$-MoTe$_2$, and a recent calculation predicts the emergence of a CDW phase, but only in the highly electron-doped regime, which lies well beyond the carrier density range explored in this study and in previous works \cite{Lee_2021, Jindal, Li_Rhodes_2024, Rhodes, Tang_2023}. Therefore, an alternative mechanism, unrelated to CDW suppression, must be responsible for the $T_c$ enhancement observed in few-layer $T_{\rm d}$-MoTe$_2$. Identifying this mechanism remains an open and compelling direction for future investigation.

\begin{acknowledgments}
We gratefully acknowledge M. Imai, S. Sasaki, and S. Wang for their support in the experiments. This project is financially supported in part by KAKENHI (Grant Numbers JP23H04869, JP21H01022, JP21H01041, JP21H04652, JP21H05236, JP21K18181, JP20H00354 and JP19H05790). K. W. and T. T. acknowledge support from the JSPS KAKENHI (Grand Numbers 21H05233 and 23H02052), the CREST \\
(JPMJCR24A5), JST and World Premier International Research Center Initiative (WPI), MEXT, Japan. 
\end{acknowledgments} 

\appendix

\section{Carrier density estimation via the normal state magnetotransport measurements}
\begin{figure*}[tb!]
\begin{center}
\includegraphics[width=14cm,clip]{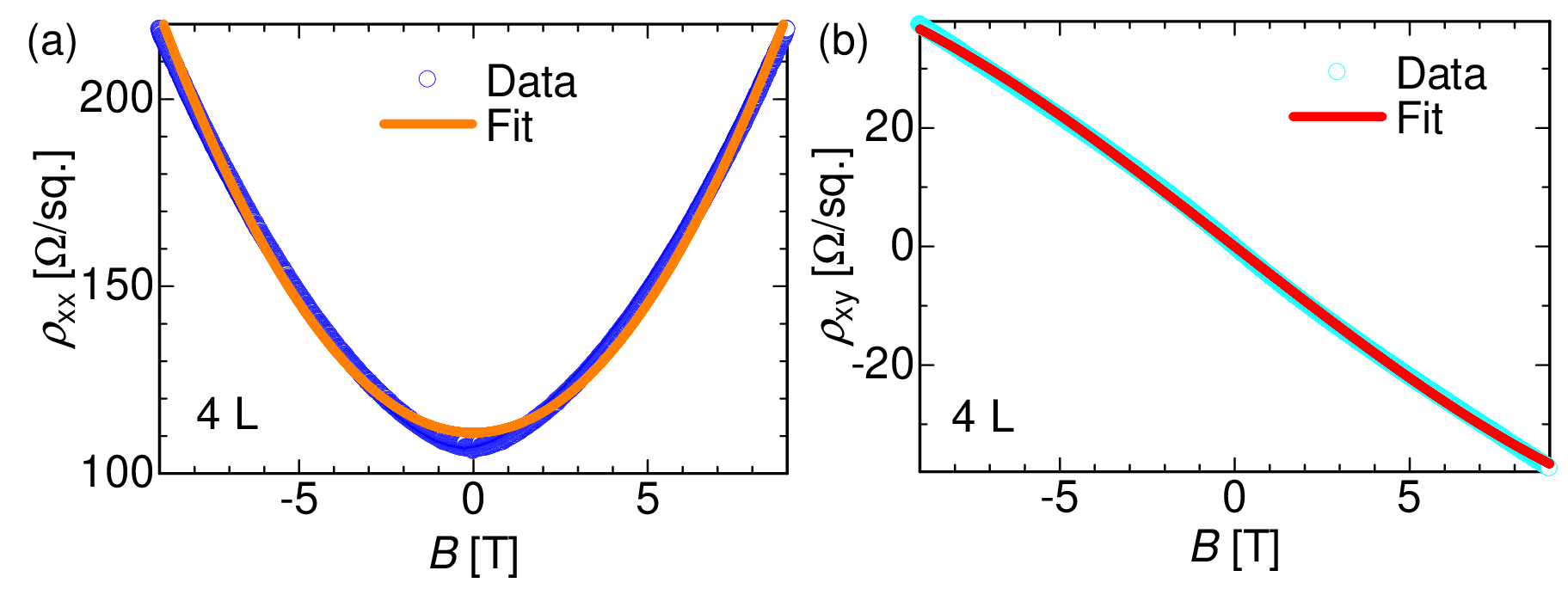}
\caption[S1]{(a) $\rho_{xx}$ from a 4 L sample (MTS14). (b) $\rho_{xy}$ simultaneously measured with $\rho_{xx}$ in (a). Solid lines are the fits based on (A1) ((a)) or (A2) ((b)).}
\label{figS1}
\end{center}
\end{figure*}   

To investigate the relationship between superconducting properties and carrier densities or mobilities, we estimate these quantities from magnetotransport measurements performed in the normal state. These estimates also allow us to determine the Fermi level of each sample, which is important for evaluating the possible pairing symmetry of the superconducting state, as discussed in the main text. Considering the semimetallic band structure of $T_{\rm d}$-MoTe$_2$, we adopt a two-carrier model that describes the longitudinal magnetoresistivity and Hall resistivity under a perpendicular magnetic field $B$ as follows: \cite{formula}
\begin{gather}
\rho_{xx}=\frac{(n_e \mu_e + n_h \mu_h)+(n_e \mu_e \mu_h^2 + n_h \mu_e^2 \mu_h) B^2}{e[(n_e \mu_e + n_h \mu_h)^2+(n_h-n_e)^2 \mu_e^2 \mu_h^2 B^2]}\\
\rho_{xy}=\frac{(n_h \mu_h^2 - n_e \mu_e^2) B + \mu_e^2 \mu_h^2 (n_h-n_e) B^3}{e[(n_e \mu_e + n_h \mu_h)^2+(n_h-n_e)^2 \mu_e^2 \mu_h^2 B^2]}
\end{gather}
where $n_e, n_h, \mu_e$, and $\mu_h$ are respectively the carrier density of electrons, that of holes, and the electron and hole mobilities. Figures \ref{figS1}(a) and (b) display the longitudinal ($\rho_{xx}$) and Hall ($\rho_{xy}$) resistivity as a function of $B$ from one of the 4 L samples measured at the temperature ($T$) = 4 K. We simultaneously fit the experimental data using equations (A1) and (A2), and the fitted results are shown as solid lines in the corresponding figures. From the fits, we obtained $n_e$ = 2.401$\pm$0.003 $\times$ 10$^{13}$ cm$^{-2}$, $n_h$ = 2.616 $\pm$0.005 $\times$ 10$^{13}$ cm$^{-2}$, $\mu_e$ = 1363$\pm$1 cm$^2$V$^{-1}$s$^{-1}$, and $\mu_h$ = 901$\pm$1 cm$^2$V$^{-1}$s$^{-1}$. Note that, for some samples, only longitudinal resistance was measured due to the limited flake size. The fitting parameters ($n_e$, $n_h$, $\mu_e$, and $\mu_h$) obtained for samples with different thicknesses are discussed in the main text in relation to the variation of $T_c$.

\section{Details of the theoretical calculations}
\label{sec:details_theory}

The DFT calculations for bulk, 2 L, and 4 L $T_{\rm d}$ MoTe$_2$ are performed using \textsc{Quantum ESPRESSO}~\cite{Giannozzi2017}.
We use the fully relativistic projector augmented-wave (PAW) pseudopotentials from \texttt{pslibrary} (ver.~1.0.0)~\cite{DalCorso2014}, where the generalized gradient approximation with Perdew-Burke-Ernzerhof parameterization for solids (PBEsol)~\cite{Perdew2008} is employed for exchange-correlation functional. 
We use 8$\times$6$\times$4 $\bm{k}$-point sampling for bulk $T_{\rm d}$-MoTe$_2$ and 
8$\times$6$\times$2 $\bm{k}$-point sampling for 2 L and 4 L $T_{\rm d}$-MoTe$_2$.
The energy cutoff of the plane-wave basis is set to 90 Ry for wave functions, and 360 Ry for the charge density. 

For 2 L and 4 L $T_{\rm d}$-MoTe$_2$, we perform structural optimization using a supercell including a vacuum region. The in-plane lattice constants are fixed to experimental values, $a= 3.477 $ \AA \ and $b = 6.335$ \AA~\cite{Qi}. The out-of-plane lattice constant is set to $c= 27.778$ \AA \ (twice the experimental lattice constant $c = 13.889$ \AA~\cite{Qi}) for the 2 L system, and  $c = 41.667$ \AA (triple the experimental lattice constant) for the 4 L system. Namely, a vacuum region with a thickness corresponding to the out-of-plane lattice constant of $T_{\rm d}$-MoTe$_2$ is inserted in both cases.

To make a fair comparison between thin-film and bulk systems, for bulk $T_{\rm d}$-MoTe$_2$, the out-of-plane lattice constant as well as the internal coordinates is relaxed, whereas the in-plane lattice constants are fixed to the experimental values.  
The obtained out-of-plane lattice constant is $c = 13.633$ \AA, in good agreement with the experimental value c = 13.889 \AA \ (1.8 \% smaller).

We then perform the band structure calculations. 
For computing the density of states, Wannier functions~\cite{Marzari1997, Souza2001} are constructed for the Mo $d$ and Te $p$ manifold by using Wannier90~\cite{Pizzi2020}. 

\nocite{*}

\bibliography{prb2}

\end{document}